\documentclass[letterpaper, 10 pt, conference]{ieeeconf}  % Comment this line out
\IEEEoverridecommandlockouts                              % This command is only
\usepackage{graphicx}
\usepackage{CJK}
\usepackage{indentfirst}
\usepackage{amsmath}
\usepackage{amssymb}
\usepackage{mathtools}
\usepackage{tabu}
\usepackage{cite}
\title{\LARGE \bf
Collision Avoidance in V2X Communication Networks
}

\author{Junwei Zang, Vahid Towhidlou and Mohammad Shikh-Bahaei% <-this % stops a space
}

\begin{document}

\maketitle
\thispagestyle{empty}
\pagestyle{empty}

%%%%%%%%%%%%%%%%%%%%%%%%%%%%%%%%%%%%%%%%%%%%%%%%%%%%%%%%%%%%%%%%%%%%%%%%%%%%%%%%
\begin{abstract}

In this paper we investigate collision detection and avoidance in a vehicular network of full duplex (FD) operating nodes. Each vehicle in this network senses the energy level of the channel before and during its transmission. The measured energy is compared against a dynamic threshold which is preset based on the target detection probability, transmitter's power, sensing time and self-interference cancellation (SIC) capability of the vehicles' on board units (OBU). Probabilities of detection and false alarm, detection threshold before and during transmission, and effect of residual self interference (SI) on these metrics have been formulated. It is shown that the proposed scheme would experience a shorter collision duration. Meanwhile, it also requires a minimum SIC capability for acceptable operation, below which, system throughput would be poor due to high false alarm probability. Numerical simulations verify the accuracy of our analysis. They also illustrate that the proposed model perform better than the fixed threshold strategy. A trade-off between half duplex (HD) and FD has been found and the scheme would be applicable even if SIC capability of OBUs is relatively poor, with no need for complicated and expensive devices for future vehicular communication.

Index Terms - full duplex, collision detection and avoidance, vehicular communication.

\end{abstract}

%%%%%%%%%%%%%%%%%%%%%%%%%%%%%%%%%%%%%%%%%%%%%%%%%%%%%%%%%%%%%%%%%%%%%%%%%%%%%%%%
\section{INTRODUCTION}
\setlength\parskip{.1\baselineskip}
Over the past few years intelligent transportation systems and smart driving have attracted the attention of auto makers and academia towards introduction of vehicular communication systems. In such networks, information such as safety messages are exchanged between vehicles (vehicle-to-vehicle V2V) or between vehicles and any other object (vehicle-to-everything V2X) to provide a better transportation system in terms of safety, latency, and energy efficiency. Two different V2X communication technologies are widely considered as promising applicants for future vehicular networks. One of them is dedicated short range communication (DSRC) \cite{c1}, which is also known as IEEE 802.11p (WAVE). While the other is cellular-V2X communication (C-V2X) such as release-15 published by 3GPP \cite{c2}. Many works have compared performance of these two technologies such as \cite{c3}. However, which technology would be the better solution is still an open question, since each of them has its own pros and cons. This work is built upon DSRC. C-V2X will be considered in our future work.

One of the main challenges based on IEEE 802.11 legacy standard (especially ad hoc networks) is the access protocol and prevention of data loss due to collision of concurrent transmission of two or many nodes. In this standard, carrier sense multiple access with collision avoidance (CSMA/CA) protocol has been deployed to minimise the probability of collision and data loss. However, this protocol will not eliminate such incidents and the condition becomes worse when there are too many nodes in the network (dense networks) \cite{c4}. This problem is more serious in vehicular networks in which safety messages known as cooperative awareness messages (CAMs) are broadcasted without acknowledgement and loss of them may result in a higher risk of accidents. Finding a way to eliminate the loss of data due to signal collision in V2X networks is an important problem to be solved.

Deploying full duplex technology in vehicular networks seems a promising solution to this problem. FD technology enables the nodes in a V2X network to sense the carrier while they are transmitting at the same time over the same channel. Thanks to recent advances in self-interference cancellation (SIC) techniques, SI suppression as high as 110 dB could be obtained under certain conditions \cite{c5}. Therefore, deploying FD technology in legacy CSMA/CA protocol enables collision detection (CD) so that vehicles would be able to detect probable collisions while broadcasting and go to backoff process in an earlier phase.

In this work we have considered collision detection and avoidance in V2X networks where the transmitting nodes would sense the channel, in a FD manner, to detect any probable concurrent transmissions. Sensing is carried out through measuring the energy level of the channel, which is a simple and widely-applied method. To the best of our knowledge, we are among the first ones exploring this method in vehicular networks. Other related works are either not in the area of vehicular networks such as \cite{c6}, \cite{c7}, \cite{c8}, \cite{c9}, \cite{c10}, \cite{c11} and \cite{c12}, or in vehicular communications but assumed ideal energy detection such as \cite{c13} and \cite{c14}. Unlike these works, we have assumed imperfect energy detection and investigated the effect of threshold, transmit power, SIC capability, sensing time, collision duration and throughput on the network performance.

The remainder of the paper is structured as follows. Section II describes the system model including assumptions of the analysis and important notations. Corresponding mathematical analysis are demonstrated in Section III. On the basis of the mathematical analysis, numerical simulations are conducted and detailed in Section IV. Finally, the paper is concluded in Section V with a brief description of our future work.

\section{SYSTEM MODEL AND NOTATIONS}

We consider a vehicular ad hoc network (VANET) in which vehicles broadcast CAMs periodically. All vehicles are equipped with FD capability. Rayleigh fading channels are assumed to be the channel model between vehicles. The noise component is assumed to be Gaussian, independent and identically distributed (i.i.d.) with zero mean and unit variance. Regarding the hidden node problem, the proposed method is not focusing on eliminating it. However, the effect of the hidden node problem could be weakened by setting up a wider sensing range such as double of the transmission range. This could be done by adjusting the thresholds formulated in section III. Since this is not the focus of this work, the transmit power level of all vehicles would be set to the same and fixed value based on the required coverage area of the vehicular network, and the sensing range is considered to be equal to the transmission range.

\begin{figure}[ht]
    \centering
    \includegraphics[width=0.45\textwidth]{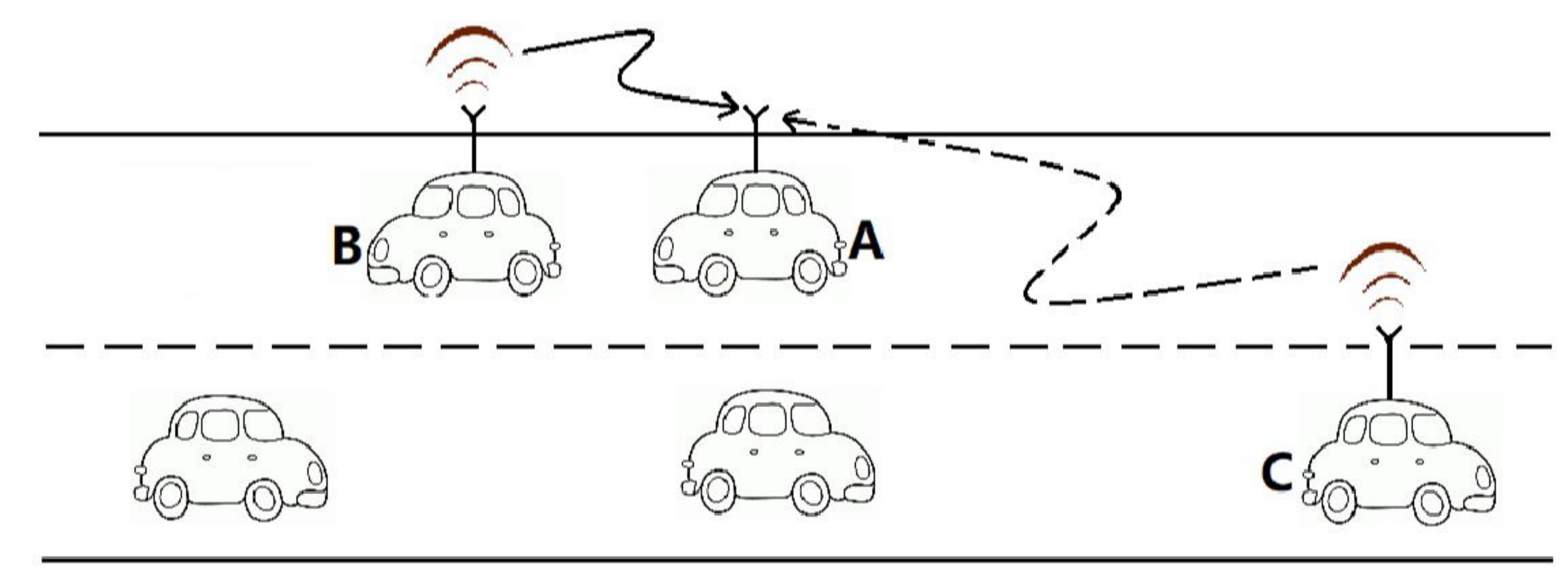}
    \caption{Demonstration of the signal detection between vehicles}
    \label{fig1}
\end{figure}

In our model, energy detection is the core technology and the level of the received energy depends on the distance between the sensing vehicle and the potential concurrent transmitting vehicle(s). To be conservative, we have developed our model to be able to detect the signal collision from the farmost vehicle, as shown in Fig. \ref{fig1}. All vehicles are able to transmit and sense simultaneously. Here vehicle A is assumed to be a transmitting and sensing vehicle, vehicle B and C are probable concurrent transmitting vehicles. It is obvious that A can easily detect the transmission of B if B is also broadcasting since B is close to A, and the measured SNR of B at A is relatively high. However, C is far away from A. If C is competing with A for broadcasting, the difficulty of detecting its transmission would be much higher than detecting B. So, we set our thresholds to satisfy the detection of signals sent by C (farmost vehicle). Certain requirements for the received SNR, sensing time and SIC capability are found, which will be discussed in later sections. Furthermore, our method performs even better when multiple CAMs are competing for broadcasting at the same time, because the energy level of the received signal would be much higher and the colliding signal is much easier to be accurately detected, comparing to the case where there is only one concurrent broadcasting vehicle.

The system works as follows, whenever a vehicle would like to broadcast a CAM, it first probes the medium to measure the energy level of the channel. If the measured energy is less than the threshold $\epsilon_{th_0}$ which is derived from the theoretical formulation in Section III, the vehicle knows the channel is free for broadcasting; if the sensed energy is greater than the threshold $\epsilon_{th_0}$, the vehicle knows there is another or even more vehicles occupying the channel and will not broadcast until the channel is free. Sensing process continues during the broadcasting period in a FD manner. The measured energy would be compared to an elevated threshold $\epsilon_{th_1}$ which is dependent on the amount of residual SI after cancellation. If the measured energy is higher than this elevated threshold $\epsilon_{th_1}$, the vehicle knows its transmission is in collision with another vehicle(s) transmitting at the same time. Otherwise, the vehicle itself is regarded as the only one using the channel in the network.

\begin{table}[ht]
  \begin{center}
    \caption{Important Notations}
    \label{tab:table1}
    \begin{tabular}{|c|l|}
    \hline
      \textbf{Parameters} & \textbf{Notes}\\
      
      \hline
      N & number of samples \\ & (maximum integer that is smaller than or equal to $\tau \cdot f_s$) \\
      \hline
      r[n] & the received signal at a FD node, \\ & where n = 1, 2, ..., N \\
      \hline
      $\tau$ & sensing time \\
      \hline
      $f_s$ & sampling frequency \\
      \hline
      $w[n]$ & noise signal with mean zero and variance $\sigma_w^2$ \\
      \hline
      $s_i[n]$ & SI signal with mean zero and variance $\sigma_i^2$ \\
      \hline
      $s[n]$ & transmit signal from a FD node \\ & with mean zero and variance $\sigma_s^2$ \\
      \hline
      $\eta$ & SIC factor \\
      \hline
      $E|.|$ & Expectation operator \\
      \hline
      $\sigma_w^2$ & variance of $w[n]$ ($\sigma_w^2 = E|w[n]|^2$) \\
      \hline
      $\sigma_i^2$ & variance of $s_i[n]$ ($\sigma_i^2 = E|s_i[n]|^2$) \\
      \hline
      $\sigma_s^2$ & variance of $s[n]$ ($\sigma_s^2 = E|s[n]|^2$) \\
      \hline
      E & energy detection test statistic \\
      \hline
      $\Upsilon_1$ & measured SNR of the node itself ($\Upsilon_1 = \frac{\sigma_i^2}{\sigma_w^2}$) \\
      \hline
      $\Upsilon_2$ & measured SNR of other transmitting node ($\Upsilon_2 = \frac{\sigma_s^2}{\sigma_w^2}$) \\
      \hline
      $\epsilon_{th_0}$ & first threshold \\
      \hline
      $\epsilon_{th_1}$ & second threshold \\
      \hline
      $H_i$ & hypothesis i where i = 0, 1, 2, 3 \\
      \hline
      $P_{f,bt}$ & probability of false alarm before transmission \\
      \hline
      $P_{f,dt}$ & probability of false alarm during transmission \\
      \hline
      $P_{d,bt}$ & probability of detection before transmission \\
      \hline
      $P_{d,dt}$ & probability of detection during transmission \\
      \hline
      $Q(.)$ & Q function operation \\
      \hline
      $p_i(x)$ & PDF of E under hypothesis $H_i$ \\
      \hline
      $\mu_i$ & mean value of $p_i(x)$ \\
      \hline
      $\sigma_i^2$ & variance of $p_i(x)$ \\
      \hline
    \end{tabular}
    \end{center}
    \vspace{-5mm}  
\end{table}
\vspace{+2mm}

However, the aforementioned detection is not perfect. All decisions are made with certain probabilities. Detection probability is defined as the probability that a vehicle successfully detect the presence of an event (an ongoing transmission or a collision) when the event actually exists, and false alarm probability is defined as the the probability that a vehicle falsely declare the presence of an event when the event does not exist. In order to have a high probability of detection, both thresholds ($\epsilon_{th_0}$, $\epsilon_{th_1}$) should be set to a low value. However, this will result in a high false alarm probability. In other words, we are missing opportunities to transmit. In this paper, we focus on keeping the detection probability to a high value while attempting to find the requirements for an acceptable probability of false alarm in terms of transmission power, sensing duration and SIC capability. In this paper we just focus on detecting collisions of signals, and further actions to be followed when collision or false alarm occurs will be left to a MAC layer scheduling protocol considered in our future work.

In order to make the mathematical formulations clear, we list the important notations in Table. \ref{tab:table1}. Specifically, $\eta$ refers to SIC factor which is the percentage of residual SI after SIC and it varies between 0 and 1. If $\eta = 0$, it means that SIC is perfect and there is no residual SI.

\section{MATHEMATICAL ANALYSIS}

From now on when we refer to transmitting vehicle we mean the vehicle which is or going to transmit and sense the channel, and colliding vehicle(s) means that the vehicle(s) that are causing collision due to concurrent transmission.

We have four different hypotheses for different transmission scenarios. Hypothesis $H_0$ is defined as when there is no vehicles broadcasting; $H_1$ is defined as when there is an ongoing transmission from colliding vehicle(s); $H_2$ is defined as when the transmitting vehicle is the only vehicle occupying the channel and $H_3$ is defined as when there are at least 2 vehicles competing for broadcasting.

So the received signal at a FD-enabled vehicle would be
\vspace{-1mm}
\begin{equation}
    \begin{array}{l}
        r[n]=\left\{
            \begin{aligned}
                &w[n]; \quad H_0
                \\
                &s[n]+w[n]; \quad H_1
                \\
                &\sqrt{\eta}s_i[n] + w[n]; \quad H_2
                \\
                &\sqrt{\eta}s_i[n] + s[n] + w[n]; \quad \ H_3
            \end{aligned}
        \right.
    \end{array}
\end{equation}

The energy detection test statistic is given by
\begin{equation}
    \begin{array}{l}
        E = \frac{1}{N}\cdot\sum_{n=1}^{N}|r[n]|^2
    \end{array}
\end{equation}

\begin{itemize}

    \item Under $H_0$:
    
    $E$ is a random variable (RV) whose probability density function (PDF) $p_0(x)$ follows a Chi-square distribution for the complex-valued case, therefore, probability of false alarm can be expressed as \cite{c6}
    
    \begin{equation}
        \begin{array}{l}
            P_{f,bt}(FD) =  Q((\frac{\epsilon_{th_0}}{\sigma_w^2}-1)\cdot\sqrt{N})
        \end{array}
    \end{equation}
    
    \item Under $H_1$:
    
    Probability of detection under this hypothesis is given by \cite{c6}
    \begin{equation}
        \begin{array}{l}
            P_{d,bt}(FD) =  Q((\frac{\epsilon_{th_0}}{\sigma_w^2}-\Upsilon_2-1)\cdot\sqrt{\frac{N}{2\Upsilon_2+1}})
        \end{array}
    \end{equation}
    
    \item Under $H_2$:
    
    Similar to $H_1$, probability of false alarm is derived from
    \begin{equation}
        \begin{array}{l}
            P_{f,dt}(FD) = Pr\{E > \epsilon_{th_1} | H_2\} = \int_{\epsilon_{th_0}}^{\infty}p_2(x)dx
        \end{array}
    \end{equation}
    
    According to Central Limit Theorem (CLT), for a large N, $p_2(x)$ can be approximated by a Gaussian distribution with mean 
     $\mu_2 = \eta^2\sigma_i^2 + \sigma_w^2$ and variance 
     $\sigma_2^2 = \frac{1}{N}\cdot[\eta^4\sigma_i^4 + 4\sigma_w^2 - (\eta^2\sigma_i^2 - \sigma_w^2)^2]$
    
    Then PDF of the measured energy would be
    \begin{equation}
        \begin{array}{l}
            p_2(x) = \frac{1}{\sqrt{2\pi\sigma_2^2}}\cdot exp(-\frac{(x-\mu_2)^2}{2\sigma_2^2})
        \end{array}
    \end{equation}
    
    Therefore, probability of false alarm is given by
    \begin{equation}
        \begin{array}{l}
            P_{f,dt}(FD) = \int_{\epsilon_{th_1}}^{\infty}\frac{1}{\sqrt{2\pi}}\cdot\frac{1}{\sqrt{\frac{1}{N}\cdot(2\eta^2\sigma_i^2\sigma_w^2+\sigma_w^4)}}\cdot \\ exp(-\frac{[x-(\eta^2\sigma_i^2+\sigma_w^2)]^2}{2[\frac{1}{N}\cdot(2\eta^2\sigma_i^2\sigma_w^2+\sigma_w^2)]})dx \\
        \end{array}
    \end{equation}
    
    After simplification, we can represent $P_{f,dt}(FD)$ in terms of the Q function as
    \begin{equation}
        \begin{array}{l}
            P_{f,dt}(FD) = Q((\frac{\epsilon_{th_1}}{\sigma_w^2} - \eta^2\Upsilon_1 - 1)\cdot\sqrt{\frac{N}{2\eta^2\Upsilon_1 + 1}})
        \end{array}
    \end{equation}
    
    \item Under H3:
    
    Similar to the previous hypotheses, probability of detection during transmission is given by
    \begin{equation}
        \begin{array}{l}
            P_{d,dt}(FD) = Pr\{E > \epsilon_{th_1} | H3 \} = \int_{\epsilon_{th_1}}^{\infty}p_3(x)dx
        \end{array}
    \end{equation}
    
    $p_3(x)$ can be approximated by a Gaussian distribution with mean $\mu_3 = \sigma_s^2 + \eta^2\sigma_i^2 + \sigma_w^2$ and variance $\sigma_3^2 = \frac{1}{N}\cdot(2\eta^2\sigma_i^2\sigma_w^2 + 2\eta^2\sigma_i^2\sigma_s^2 + 2\sigma_s^2\sigma_w^2 + \sigma_w^4)$
    
    We use the same method to derive $P_{d,dt}(FD)$ as
    \begin{equation}
        \begin{array}{l}
            P_{d,dt}(FD) = Q((\frac{\epsilon_{th_1}}{\sigma_w^2}-\Upsilon_2-\eta^2\Upsilon_1-1)\\ \cdot \sqrt{\frac{N}{2\eta^2\Upsilon_1+2\eta^2\Upsilon_1\Upsilon_2+2\Upsilon_2+1}}) \\
        \end{array}
    \end{equation}
    
    \item Thresholds $\epsilon_{th_0}$ \& $\epsilon_{th_1}$:
    
    Threshold $\epsilon_{th_0}$ is found from (4) by calculating the inverse Q function, which is given by
    \begin{equation}
        \begin{array}{l}
            \epsilon_{th_0} = (\frac{(Q^{-1}(\overline{P_{d,bt}(FD)})}{\sqrt{\frac{N}{2\Upsilon_2+1}}}+\Upsilon_2+1)\cdot\sigma_w^2
        \end{array}
    \end{equation}
    
    Threshold $\epsilon_{th_1}$ is defined from (9) as
    \begin{equation}
        \begin{array}{l}
            \epsilon_{th_1} = (\frac{(Q^{-1}(\overline{P_{d,dt}(FD)})}{\sqrt{\frac{N}{2\eta^2\Upsilon_1+2\eta^2\Upsilon_1\Upsilon_2+2\Upsilon_2+1}}}+\Upsilon_2+\eta^2\Upsilon_1+1)\cdot\sigma_w^2
        \end{array}
    \end{equation}
    
    \begin{figure}[ht]
        \centering
        \vspace{-3mm}
        \includegraphics[width=0.4\textwidth]{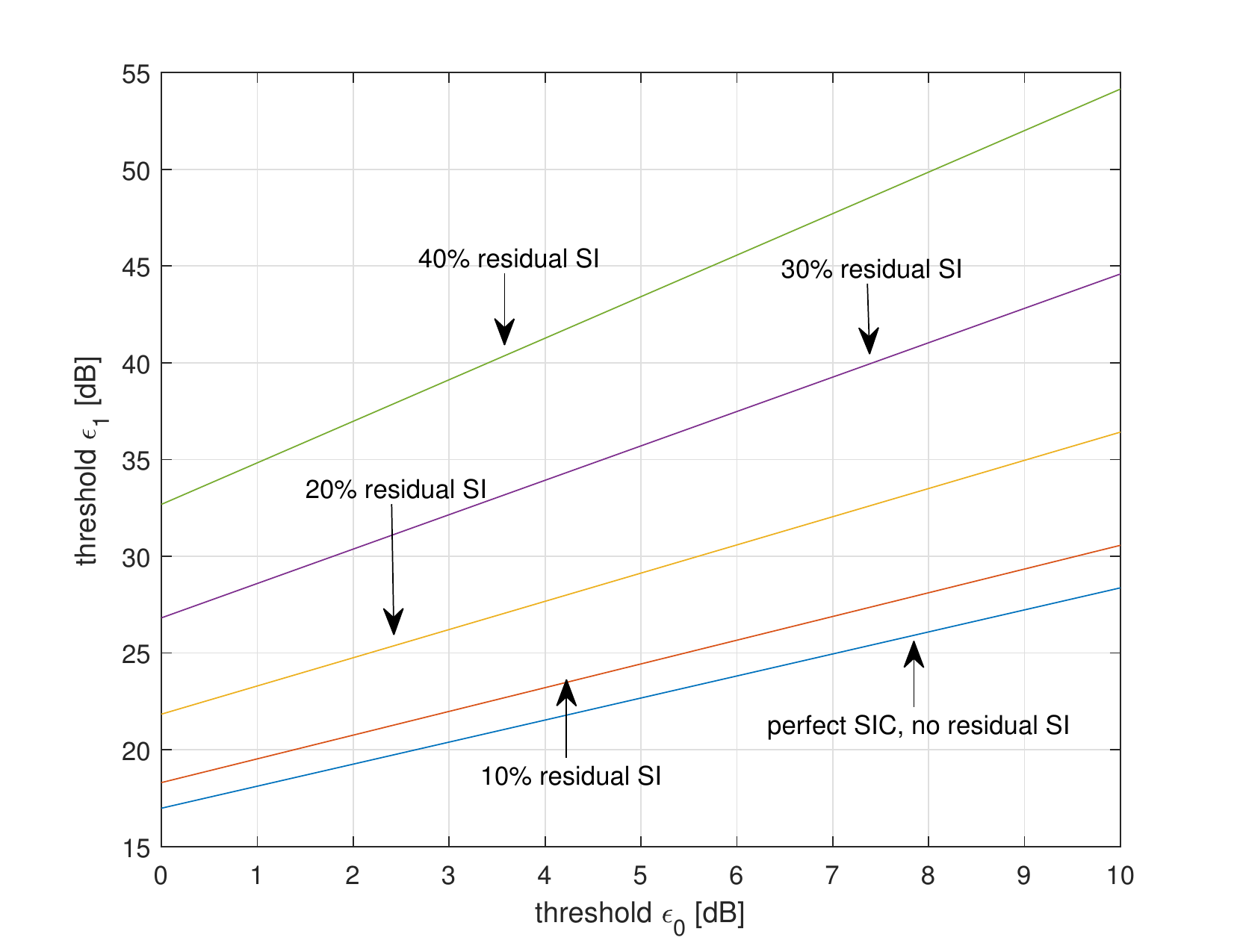}
        \vspace{-6mm}
        \caption{relationship between threshold $\epsilon_{th_0}$ and threshold $\epsilon_{th_1}$}
        \vspace{-3mm}
        \label{fig2}
    \end{figure}
    
    Assume the target probabilities of detection before and during transmission are identical, then the relationship between the two thresholds can be derived as
    \begin{equation}
        \begin{array}{l}
            \epsilon_{th_1} = \frac{\frac{\epsilon_{th_0}}{\sigma_w^2}-\Upsilon_2-1}{\sqrt{\frac{2\Upsilon_2+1}{2\eta^2\Upsilon_1+2\eta^2\Upsilon_1\Upsilon_2+2\Upsilon_2+1}}}+\eta^2\Upsilon_1+\Upsilon_2+1
        \end{array}
    \end{equation}
    
    This relationship is depicted in Fig. \ref{fig2}. It is obvious that the higher the residual SI is, the higher the thresholds would be, and the bigger the difference between the two thresholds would be.
    
    \item SIC factor $\eta$:
    
    SIC capability plays an important role in detecting collisions during transmission. With a huge amount of residual SI, both probabilities become worse. In order to see the impact of residual SI, we derive the SIC factor $\eta$ in terms of the false alarm probability as
    \begin{equation}
        \begin{array}{l}
            \eta = \frac{2N(\frac{\epsilon_{th_1}}{\sigma_w^2}-0.5)+2[Q^{-1}(\overline{P_{f,dt}(FD)})]^2-N}{2N\Upsilon_1} \\
            + \frac{\sqrt{8(\frac{\epsilon_{th_1}}{\sigma_w^2}-0.5)N[Q^{-1}(\overline{P_{f,dt}(FD)})]^2+4[Q^{-1}(\overline{P_{f,dt}(FD)})]^4}}{2N\Upsilon_1}
        \end{array}
    \end{equation}
    
    (15) has solutions only when
    
    $\Delta = \frac{8yN[Q^{-1}(\overline{P_{f,dt}(FD)})]^2+4[Q^{-1}(\overline{P_{f,dt}(FD)})]^4}{N^2}\geqslant 0$, where $y = \frac{\epsilon_{th_1}}{\sigma_w^2}-0.5$.
    
    \item Average probability of false alarm:
    
    SIC factor is not always fixed, it may fluctuate due to the imperfection of the hardware or channel variations. For a given SIC factor $\eta_0$ with $\pm m\%$ fluctuation distributed uniformly, the average probability of false alarm can be calculated by
    \begin{equation}
        \begin{array}{l}
            \overline{P_{f,dt}(FD)} = \int_{\eta_0-m}^{\eta_0+m}Q((\frac{\epsilon_{th_1}}{\sigma_w^2} - \eta_0^2\Upsilon_1 - 1)\cdot \\
            \sqrt{\frac{N}{2\eta_0^2\Upsilon_1 + 1}})\cdot PDF(\eta) d\eta
        \end{array}
    \end{equation}
    
    According to \cite{c15}, the Q function could be approximated as $Q(x) \approx \frac{1}{2}e^{-\frac{x^2}{2}}$.
    Thus, the average probability of false alarm is derived as
    \begin{equation}
        \begin{array}{l}
            \overline{P_{f,dt}(FD)} = \frac{1}{4m}\int e^{-\frac{(y-z)^2N}{4z}}\cdot\frac{1}{\Upsilon_1} dz
        \end{array}
    \end{equation}
    
    where $y = \frac{\epsilon_{th_1}}{\sigma_w^2}-0.5$ and $z = \eta_0^2\Upsilon_1+0.5$. This integral cannot be solved and the closed-form expression cannot be found. However, we can approximate the average probability of false alarm according to the simulation result shown in Fig. \ref{fig3} as an example. We found that regardless of target probability, initial SIC factor and SIC fluctuation, the average probability of false alarm is always the average of $P_f$ for $\eta_0+m$ and $\eta_0-m$.
    \begin{figure}[ht]
        \centering
        \vspace{-3mm}
        \includegraphics[width=0.4\textwidth]{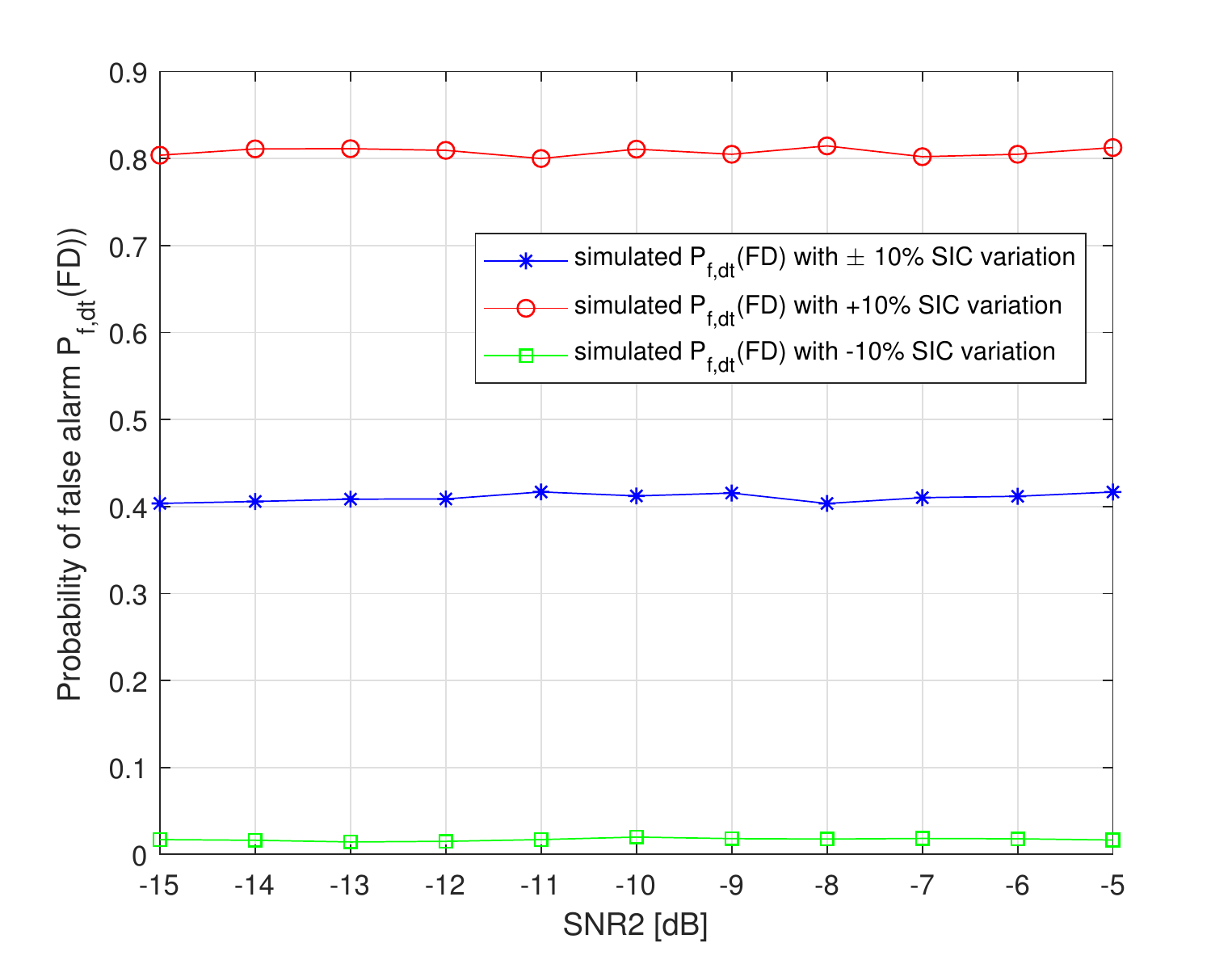}
        \vspace{-5mm}
        \caption{Probability of false alarm $P_{f,dt}(FD)$ VS SNR2 with SIC variation}
        \vspace{-3mm}
        \label{fig3}
    \end{figure}
    
    Thus the average probability of false alarm is expressed as
    \begin{equation}
        \begin{array}{l}
            \overline{P_{f,dt}(FD)} \approx \frac{Q((\frac{\epsilon_{th_1}}{\sigma_w^2} - (\eta_0+m)^2\Upsilon_1 - 1)\sqrt{\frac{N}{2(\eta_0+m)^2\Upsilon_1 + 1}})}{2}\\
            + \frac{Q((\frac{\epsilon_{th_1}}{\sigma_w^2} - (\eta_0-m)^2\Upsilon_1 - 1)\sqrt{\frac{N}{2(\eta_0+m)^2\Upsilon_1 + 1}})}{2}
        \end{array}
    \end{equation}
    
\end{itemize}

\section{SIMULATION RESULTS}

Following the mathematical analysis, we now evaluate our proposed method through simulations. Relevant simulation parameters are shown in Table \ref{tab:table2}.

\begin{table}[ht]
  \begin{center}
    \caption{Parameters and Assumptions}
    \label{tab:table2}
    \begin{tabular}{|l|r|}
    \hline
    \textbf{Parameters} & \textbf{Values}\\
      
      \hline
      target $P_{d,bt}(FD)$ \& $P_{d,dt}(FD)$ & 90\% \& 50\%\\
      \hline
      modulation scheme & BPSK, QPSK\\
      \hline
      measured SNR from SI signal before SIC (SNR1) & +10 dB\\
      \hline
      measured SNR from another vehicle (SNR2) & (-20)$\rightarrow$0 dB\\
      \hline
      Residual SI & 0\%-40\%\\
      \hline
      vehicle density & 0-200 vehicles/km\\
      \hline
    \end{tabular}
  \end{center}
  \vspace{-5mm}
\end{table}

\begin{figure}[ht]
    \centering
    \vspace{-3mm}
    \includegraphics[width=0.4\textwidth]{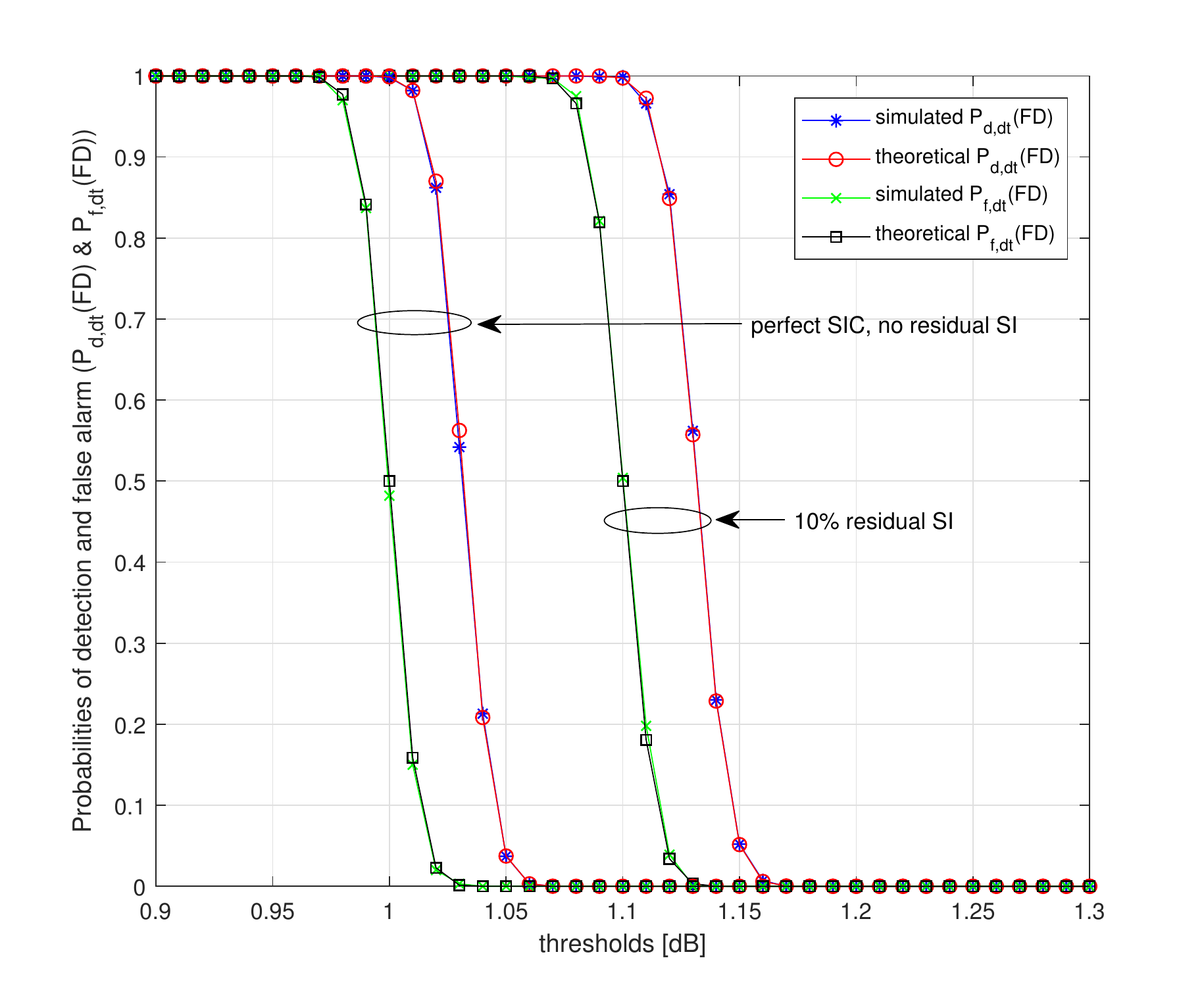}
    \vspace{-5mm}
    \caption{Probabilities of detection $P_{d,dt}(FD)$ and false alarm $P_{f,dt}(FD)$ VS threshold $\epsilon_{th_1}$ under different SIC assumptions}
    \vspace{-3mm}
    \label{fig4}
\end{figure}

Fig.\ref{fig4} shows a good match between the simulated $P_{d,dt}(FD)$ and $P_{f,dt}(FD)$ and their theoretical values, which verifies our mathematical analysis to be correct and accurate. When residual SI becomes stronger, in order to achieve the same detection probability, thresholds should be set to a higher value because more energy (from SI signal) is received. Moreover, a small variation of the threshold $\epsilon_{th_1}$ would result in a huge deviation in the probabilities even when SIC is perfect. For example,  if the threshold changes from 1 dB to 1.025 dB, such a small change would lead to 45\% drift of the probabilities of detection and false alarm. This result highlights the calculation of the threshold is of great importance and should be done as accurate as possible.

\begin{figure}[ht]
    \centering
    \includegraphics[width=0.4\textwidth]{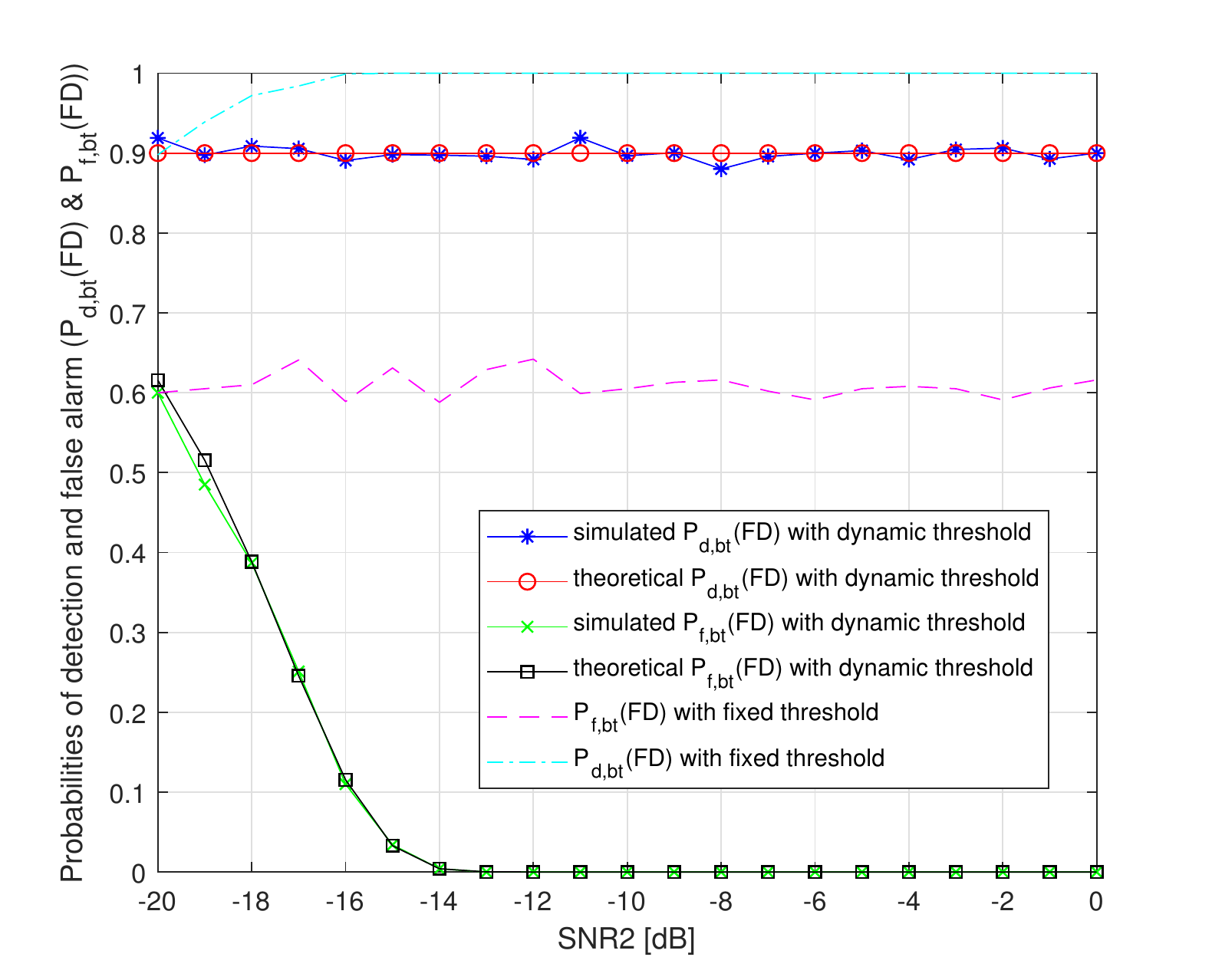}
    \vspace{-5mm}
    \caption{Probabilities of detection $P_{d,bt}(FD)$ and false alarm $P_{f,bt}(FD)$ VS measured SNR before transmission}
    \vspace{-2mm}
    \label{fig5}
\end{figure}

\begin{figure}[ht]
    \centering
    \vspace{2mm}
    \includegraphics[width=0.4\textwidth]{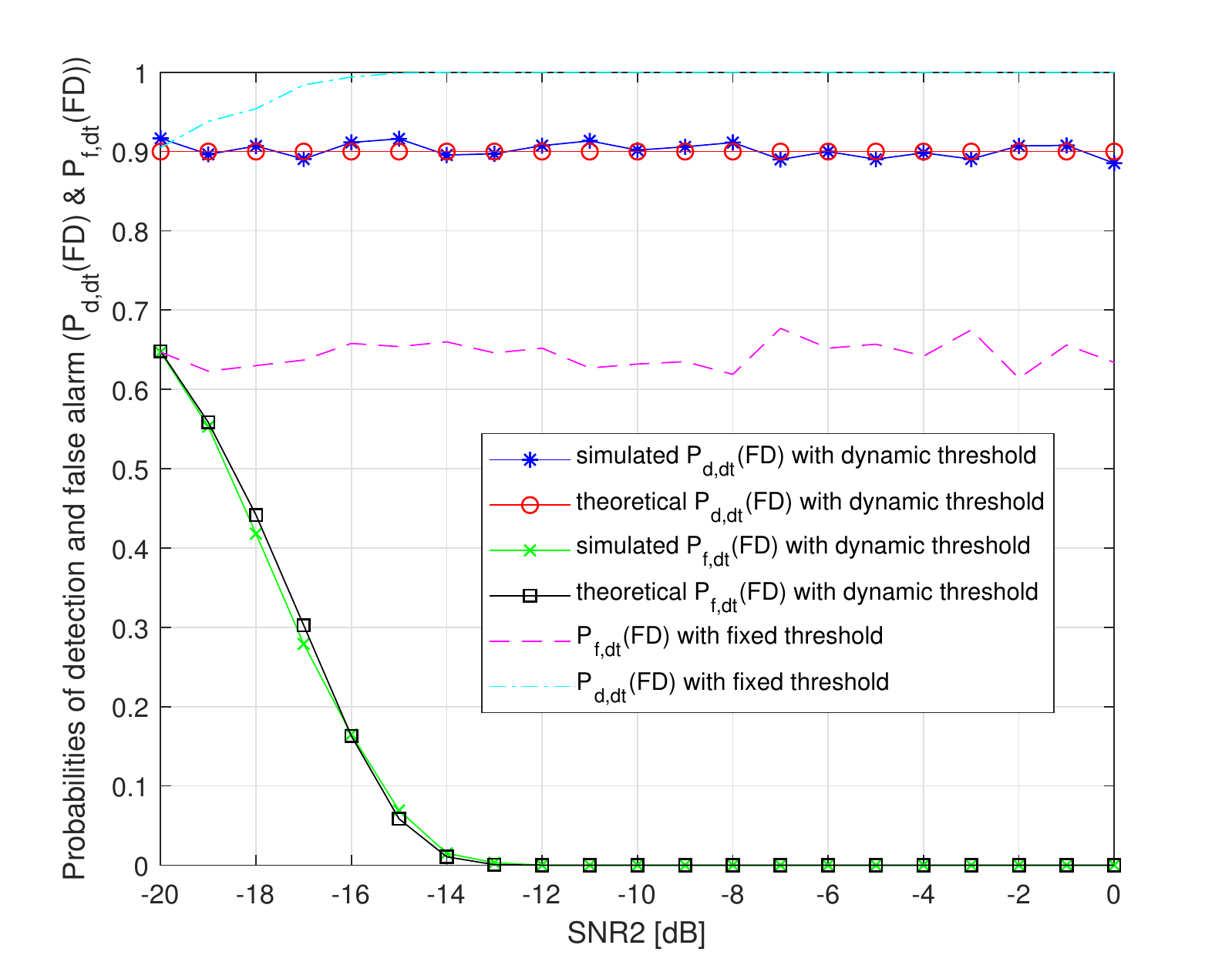}
    \vspace{-5mm}
    \caption{Probabilities of detection $P_{d,dt}(FD)$ and false alarm $P_{f,dt}(FD)$ VS measured SNR during transmission}
    \vspace{-3mm}
    \label{fig6}
\end{figure}

Fig. \ref{fig5} and Fig. \ref{fig6} illustrate the significant impact of transmit power and the difference between two threshold setting strategies. One is our proposed method where the threshold is dynamically changing, while the other is the fixed threshold method. For the fixed threshold strategy, along with the rise of the transmit power, probability of detection increases while having a high and unacceptable probability of false alarm. Our proposed method would have a lower detection probability which is still in the acceptable range. But because the threshold is increasing too with the rise of the measured SNR, false alarm probability would decrease at the same time. Compared to the fixed threshold method, although our proposed method will sacrifice some detection probability by dynamically changing the threshold, a much better false alarm probability would be rewarded, while keeping the probability of detection in an acceptable range.

\begin{figure}[ht]
    \centering
    \includegraphics[width=0.45\textwidth]{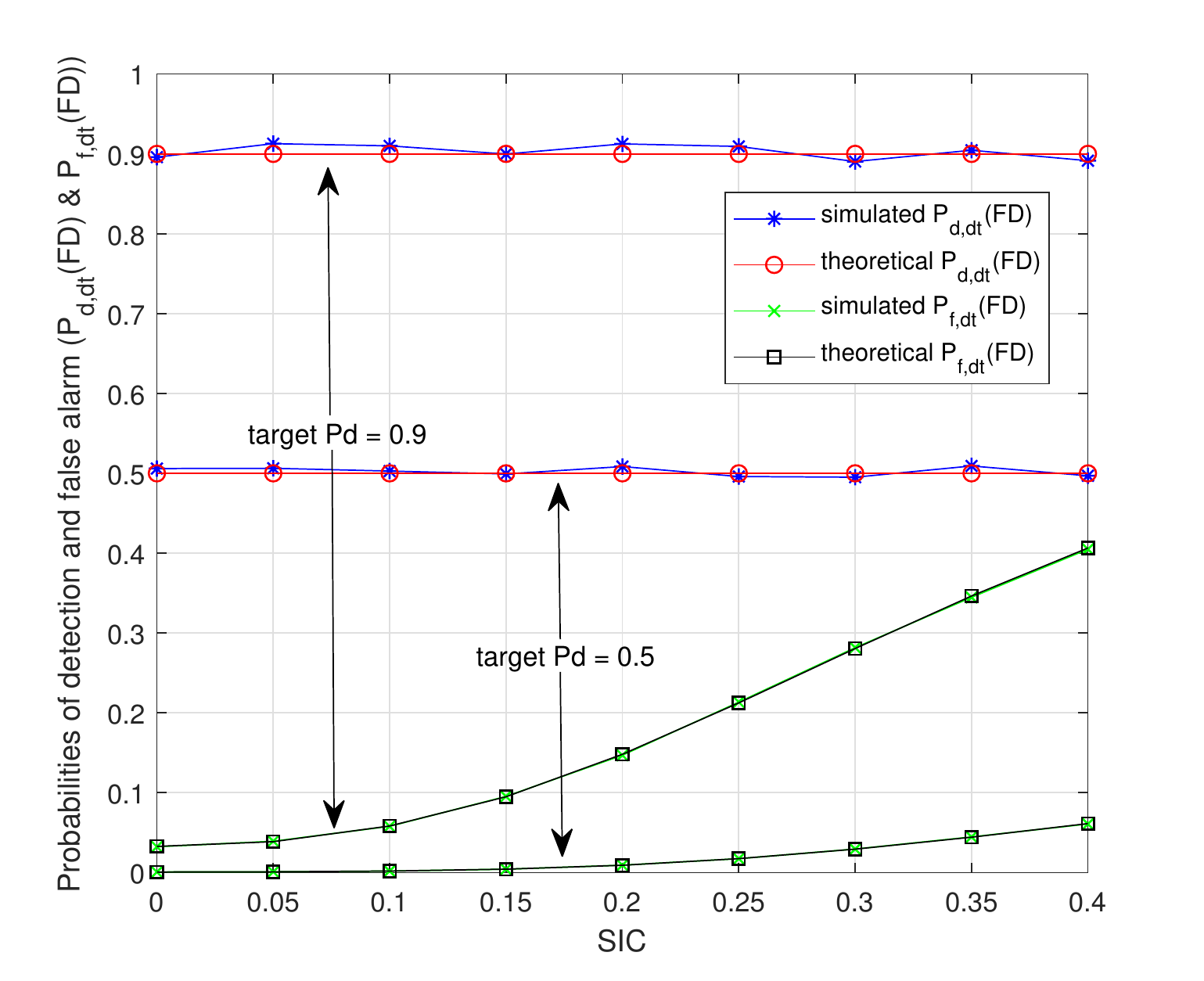}
    \vspace{-5mm}
    \caption{Probabilities of detection $P_{d,dt}(FD)$ and false alarm $P_{f,dt}(FD)$ VS SIC factor $\eta$}
    \vspace{-3mm}
    \label{fig7}
\end{figure}

Fig. \ref{fig7} shows the effect of residual SI on the probabilities. Firstly, target probability of detection is achievable regardless of SIC factor by using the dynamic threshold. However, when SIC factor $\eta$ increases, false alarm probability increases too since more energy is received. In order to achieve detection probability to be at least $90\%$ and false alarm probability to be at most $10\%$, our model would have acceptable performance when SIC factor is less than $15\%$. In other words, our system does not operate only when SIC is extremely well, it works quite acceptable when SIC is relatively poor.

\begin{figure}[ht]
    \centering
    \vspace{-2mm}
    \includegraphics[width=0.45\textwidth]{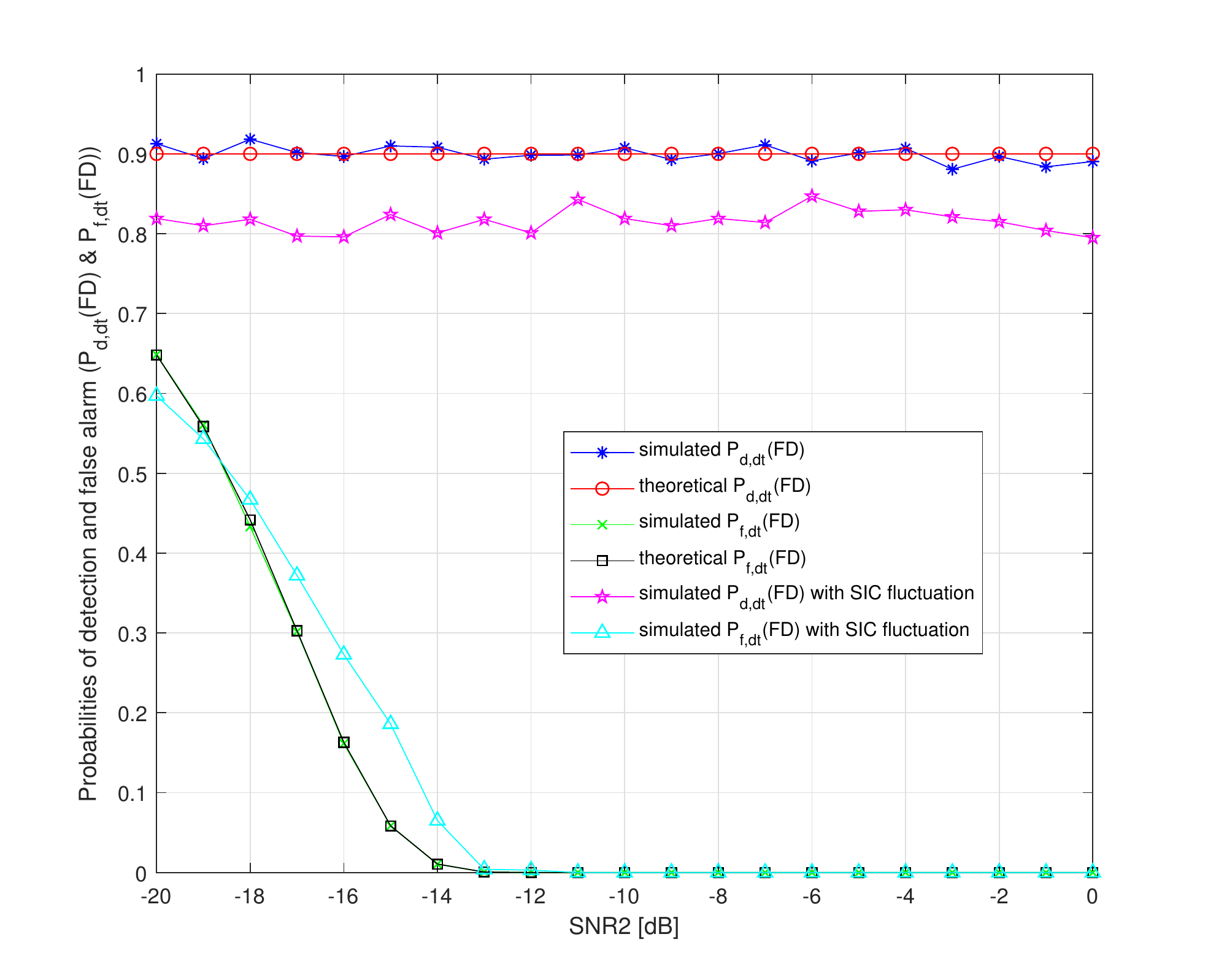}
    \vspace{-5mm}
    \caption{Probabilities of detection $P_{d,dt}(FD)$ and false alarm $P_{f,dt}(FD)$ VS measured SNR during transmission with 10\% SIC fluctuation}
    \vspace{-6mm}
    \label{fig8}
\end{figure}

Fig. \ref{fig8} highlights the damage of SIC fluctuation. When a 10\% random SIC fluctuation is considered, both probabilities of detection and false alarm become worse compared to the case where such a fluctuation does not considered. Thus, a space for SIC fluctuation should be left when deploying the scheme in future V2X systems.
\vspace{-2mm}

\begin{figure}[ht]
    \centering
    \includegraphics[width=0.4\textwidth]{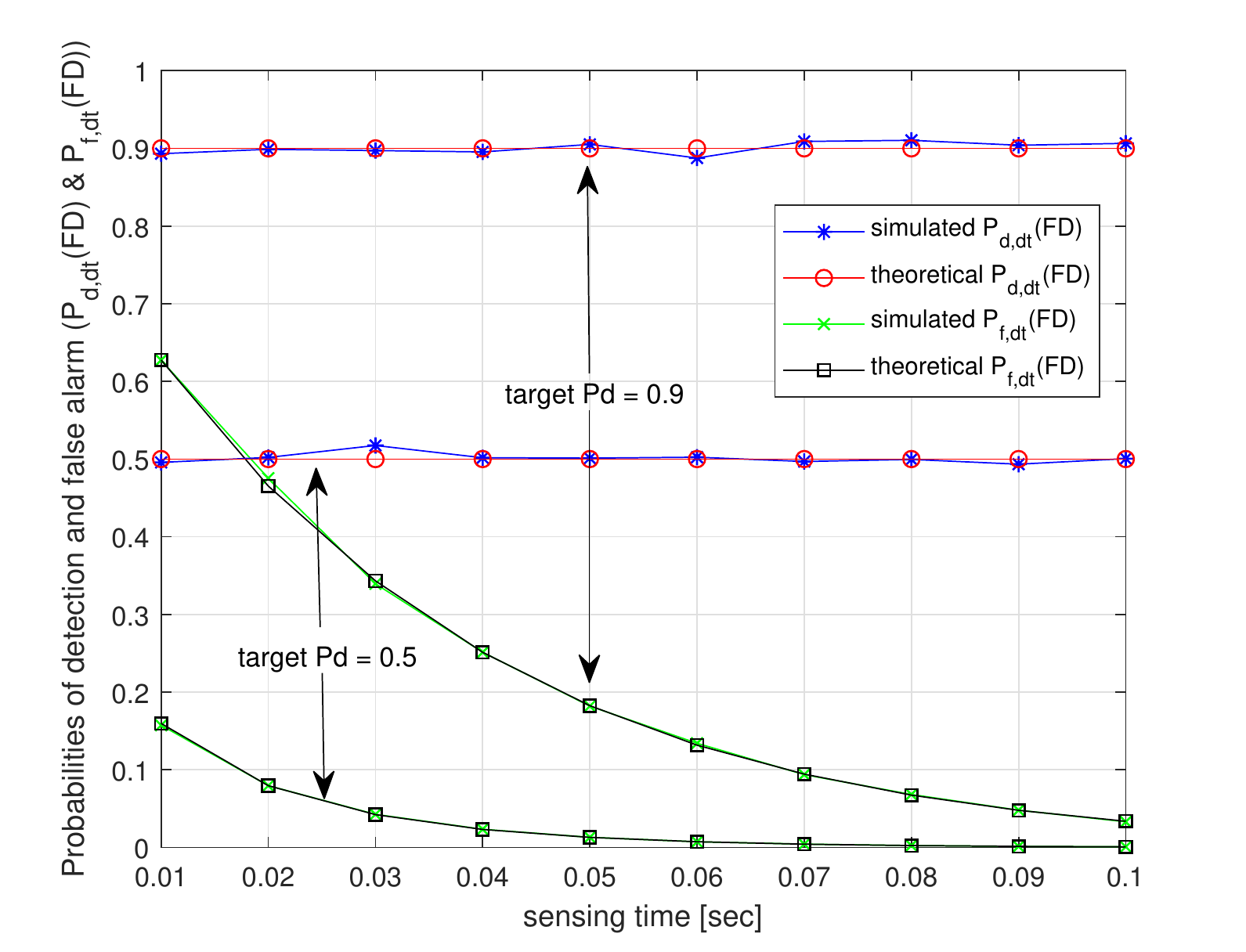}
    \vspace{-3mm}
    \caption{Probabilities of detection $P_{d,dt}(FD)$ and false alarm $P_{f,dt}(FD)$ VS sensing time during transmission}
    \vspace{-3mm}
    \label{fig9}
\end{figure}
\vspace{+2mm}
Fig. \ref{fig9} shows the impact of the sensing time on the precision of detection. By setting the thresholds properly, the system can achieve the target detection performance. Meanwhile, the longer the sensing time is, the lower chance the system would wrongly alarm an impending collision. This is because we are measuring and averaging the received energy over a longer period of time, which gives a more accurate detection result. Another way to reduce the false alarm probability is to increase the sampling frequency, since $N = \tau\cdot f_s$. However, the accuracy of the detection performance cannot be improved by only increasing $f_s$. When the number of samples taken of a signal is large enough, more samples would not give a more accurate measured energy level.

\begin{figure}[ht]
    \centering
    \includegraphics[width=0.4\textwidth]{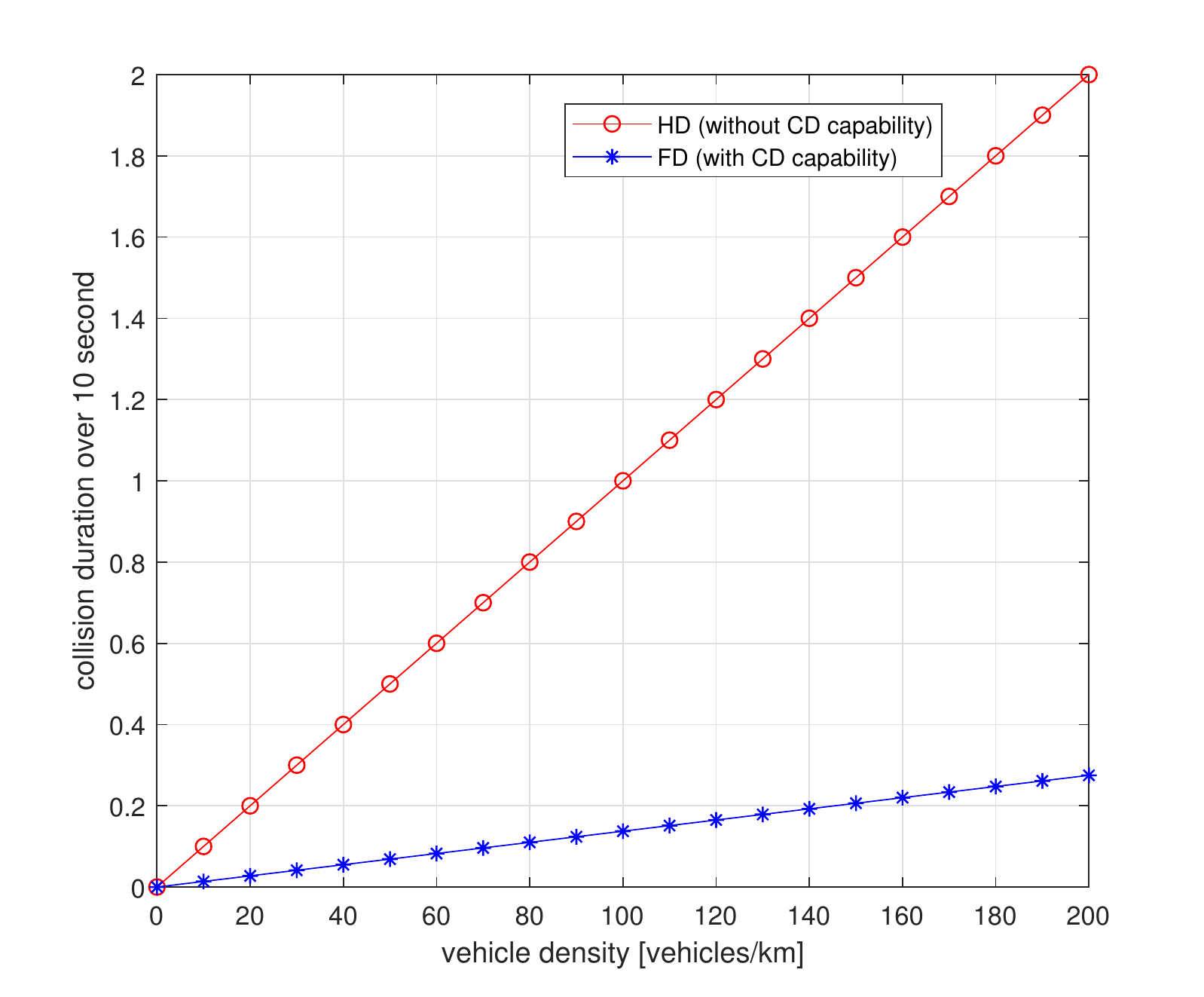}
    \vspace{-3mm}
    \caption{Collision duration over 10 seconds VS average vehicle density}
    \vspace{-3mm}
    \label{fig10}
\end{figure}

\begin{figure}[ht]
    \centering
    \includegraphics[width=0.4\textwidth]{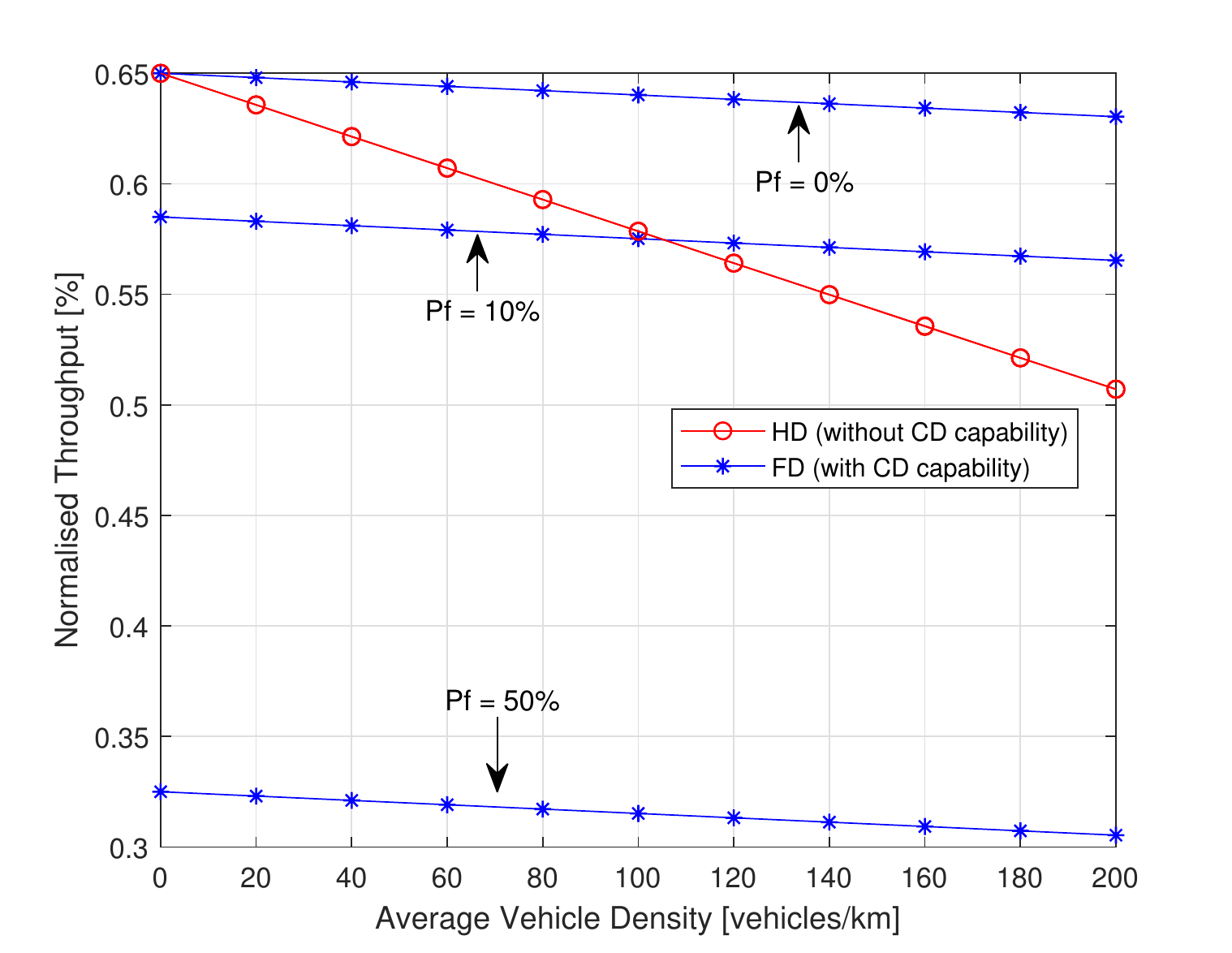}
    \vspace{-3mm}
    \caption{Normalised throughput VS average vehicle density under different probabilities of false alarm}
    \vspace{-3mm}
    \label{fig11}
    \vspace{-3mm}
\end{figure}

Fig. \ref{fig10} shows the collision duration over 10 seconds versus the average vehicle density. Vehicles are assumed to be placed on a line according to Poisson distribution, which is the same as the assumption in \cite{c13}. It is clear that vehicles with CD capability experience a shorter period of collision compared to the vehicles which operate in HD mode. The difference becomes larger when the number of vehicles increases. In other words, FD technology helps vehicles to avoid impending collisions by aborting transmissions at an earlier stage, while vehicles without CD capability would experience collision for an entire packet duration.

Fig. \ref{fig11} demonstrates the drawback of enabling CD capability. While detecting probable collisions at an early stage, we are sacrificing throughput compared to HD mode, especially when probability of false alarm is high. However, since probability of false alarm can be reduced while keeping the probability of detection in an acceptable range, throughput could be improved. Thus, a trade-off between HD and FD modes is found. When the number of vehicles in a VANET is relatively low, HD mode is preferred. FD mode is more suitable for dense VANETs.

\section{CONCLUSIONS}

In this paper we studied full-duplex collision detection and avoidance through energy detection method in V2X networks. By deploying the proposed model, a vehicle can detect and avoid collisions with certain probabilities. Two thresholds which are dynamically changing have been formulated. Simulation results have shown that our model does not require near perfect SIC, it works well even when SIC is poor. On the basis of these results, detecting and avoiding collisions in V2X communication networks could be better, more suitable MAC layer protocol based on original CSMA/CA could be proposed to provide better communication environment, which will also be our future work.

\section{ACKNOWLEDGEMENT}

This work is partially supported by EPSRC under grant number EP/P003486/1 and EC Horizon 2020 project: 5GCAR.

\addtolength{\textheight}{-12cm}   % This command serves to balance the column lengths

%%%%%%%%%%%%%%%%%%%%%%%%%%%%%%%%%%%%%%%%%%%%%%%%%%%%%%%%%%%%%%%%%%%%%%%%%%%%%%%%

%\bibliographystyle{IEEEtran}
%\bibliography{IEEEabrv,Ref}

\end{document}